\documentclass[11pt,letterpaper]{article}
\pdfoutput=1
\usepackage{a4wide,epsfig,amsmath,amssymb,cite,scalefnt,graphicx}
\numberwithin{equation}{section}
\allowdisplaybreaks 
\usepackage{array}
\usepackage{epstopdf}
\usepackage{tikz}
\usetikzlibrary{intersections}
\usetikzlibrary{calc}
\usetikzlibrary{shapes}
\usetikzlibrary{arrows,shapes,positioning}
\usetikzlibrary{decorations.markings}
\tikzstyle arrowstyle=[scale=1]
\tikzstyle directed=[postaction={decorate,decoration={markings,
    mark=at position .5 with {\arrow[arrowstyle]{stealth}}}}]
\tikzstyle reverse directed=[postaction={decorate,decoration={markings,
    mark=at position .5 with {\arrowreversed[arrowstyle]{stealth};}}}]
\newcommand{\vertex}{\node[fill,circle,inner sep=0pt,minimum size=0pt]}
\newcommand{\coord}[1]{({sin(#1)},{cos(#1)})}

\def\phibar{\bar\phi}
\def\Qbar{\bar Q}
\def\be{\begin{equation}}
\def\ee{\end{equation}}
\def\pa{\partial}
\def\Ocal{{\cal O}}
\def\sigmabar{\overline\sigma}
\def\nn{\nonumber\\}
\newcommand{\vbeta}{\node[fill=black,diamond,inner sep=3.5pt,minimum size=0pt]}

\input{epsf}

\begin{document}
%\numberwithin{equation}{section}

\begin{titlepage}
\begin{flushright}
LTH1252\\
{\today
%{\bf  Draft \jobname}
}\\

\end{flushright}
\date{}
\vspace*{3mm}

\begin{center}
{\Huge Anomalous dimensions at large charge in $d=4$ $O(N)$ theory}\\[12mm]
{\bf I.~Jack\footnote{{\tt dij@liverpool.ac.uk}} and  D.R.T.~Jones\footnote{{\tt drtj@liverpool.ac.uk}} 
}\\
%\end{center}

\vspace{5mm}
Dept. of Mathematical Sciences,
University of Liverpool, Liverpool L69 3BX, UK\\

\end{center}

\vspace{3mm}
\begin{abstract}
Recently it was shown that the scaling dimension of the operator $\phi^n$ in $\lambda(\phibar\phi)^2$ theory may be computed semiclassically at the Wilson-Fisher fixed point in $d=4-\epsilon$, for generic values of $\lambda n$, and this was verified to two loop order in perturbation theory at leading and subleading $n$. In subsequent work, this result was generalised to operators of fixed charge $\Qbar$ in $O(N)$ theory and verified up to three loops in perturbation theory at leading and subleading $\Qbar$. Here we extend this verification to four loops in $O(N)$ theory, once again at leading and subleading $\Qbar$. We also investigate the strong-coupling regime.
\end{abstract}

\vfill

\end{titlepage}

\section{Introduction}

Renormalizable theories with scale invariant scalar self-interactions have been subjects of enduring interest. In particular, the study of theories with quartic ($\phi^4$) interactions in $d=4-\epsilon$ dimensions has played a central role in the development of the theory of critical phenomena, since the pioneering work of Wilson~\cite{wils1,wils2}\ and Wilson and Fisher~\cite{wf} 
in 1971. Study of the renormalisation group flow of the coupling or couplings of the theories facilitates the determination of the order of  phase transitions and the associated critical indices. For example, the theory with a single scalar field exhibits a Wilson-Fisher fixed point (FP) where the coupling constant $\lambda$ is $O(\epsilon)$, and this infra-red (IR) attractive FP is associated with a second order phase transition. 

Historically, the majority of work in renormalisable quantum field theories has involved the weak coupling expansion, in other words the Feynman diagram loop expansion. However this expansion fails or becomes ponderous at either strong coupling or (less obviously) for $\phi^n$ amplitudes at large $n$. The latter has obviously developed in importance as collider energies have increased.  Remarkable progress~\cite{son,horm,alos,aos,rod,Bad2,sann,sann2,alos2} here came with the use of a semi-classical expansion in the path integral formulation of the theory\footnote{An analogous analysis was pursued for $\phi^6$ theories for $d=3-\epsilon$ and $\phi^3$ theories for $d=6-\epsilon$ in Refs.~\cite{Bad,rod2,JJ}}.

In Ref~\cite{rod} the anomalous dimension of the $\phi^n$ operator was considered in the  $O(N)$-invariant $g(\phi^2)^2$ theory with an $N$-dimensional scalar multiplet $\phi$, for large $n$ and  fixed $gn^2$. In Ref.~\cite{Bad2} the scaling dimension of the same operator in the $U(1)$-invariant $\lambda(\phibar\phi)^2$ theory (corresponding to the special case $N=2$) was computed at the Wilson-Fisher fixed point $\lambda_*$ as a semiclassical expansion in $\lambda_*$, for fixed $\lambda_*n$. Subsequently this was generalised in Ref.~\cite{sann} to the case of an operator of charge $\Qbar$ in the $O(N)$-invariant  theory.  In Ref.~\cite{Bad2}, the $U(1)$ result was compared with perturbation theory up to two loops, and in Ref.~\cite{sann} the check was performed for the $O(N)$ theory up to three loops. Here we proceed directly with the $O(N)$ case, since, at least for our purposes, many salient features of the analysis are very similar in both cases; and the results for $U(1)$ may be recovered from those for $O(N)$, essentially by setting $N=2$. We extend the  comparison with perturbation theory up to four loops, and also discuss the large $(g\Qbar)$ case, generalising the large $\lambda n$ analysis of Ref.~\cite{Bad2}.

The paper is organised as follows: In Section 2 we describe the semiclassical calculation in the $O(N)$ case, following Ref.~\cite{sann}. Then in Section~3 we compare the result of this calculation with perturbative calculations up to and including 4 loops. This represents a significant extension of previous calculations. In Section~\ref{gQ} we address the large $(g\Qbar)$ limit and compare in detail with earlier work.

\section{The $O(N)$ case}

In the $O(N)$ case we have a multiplet of fields $\phi_i$, $i=1\ldots N$, and the Lagrangian is 
\be
{\cal L} =\frac12\pa^{\mu}\phi_i\pa_{\mu}\phi_i+\frac{g}{4!}(\phi_i\phi_i)^2.
\ee
The $\beta$-function for this theory is well-known\cite{klein}
\be
16\pi^2\beta(g)=-\epsilon g+\frac{g^2}{3}(N+8)-\frac{g^3}{3}(3N+14)+\Ocal(g^4),
\ee
and leads to an infra-red conformal fixed point
\be
g_*=\frac{3\epsilon}{N+8}+\frac{9(3N+14)}{(N+8)^3}\epsilon^2+\Ocal(\epsilon^3).
\label{gfix}
\ee
As shown in Ref.~\cite{sann}, the fixed-charge operator of charge $\Qbar$ may be taken to be
\be
T_{\Qbar}=T_{i_1i_2\ldots i_{\Qbar}}\phi_{i_1}\phi_{i_2}\ldots \phi_{i_{\Qbar}},
\ee
where $T_{i_1i_2\ldots i_{\Qbar}}$ is symmetric,  and traceless on any pair of indices. The scaling dimension $\Delta_{T_{\Qbar}}$ is expanded as 
\be
\Delta_{T_{\Qbar}}=\Qbar\left(\frac d2-1\right)+\gamma_{T_{\Qbar}}=\sum_{\kappa=-1}g^{\kappa}\Delta_{\kappa}(g\Qbar).
\label{Tscal}
\ee
We initially work in general $d$. The semiclassical computation of $\Delta_{-1}$ and $\Delta_0$  is performed by mapping the theory via a Weyl transformation to a cylinder $\mathbb{R}\times S^{d-1}$, where $S^{d-1}$ is a sphere of radius $R$; where the ${\cal R}\phi^*\phi$ term (${\cal R}$ being the Ricci curvature) generates an effective $m^2\phi^*\phi$ mass term with $m=\frac{d-2}{2R}$. This mapping process along with other technical simplifications\cite{Bad2} relies on conformal invariance and therefore we now assume that we are at the conformal fixed point in Eq.~\eqref{gfix}. It was shown in Ref.~\cite{Bad2} that  stationary configurations of the action are characterised by a chemical potential $\mu$, related to the cylinder radius $R$ by
\be
R\mu_*=\frac{3^{\frac13}+\left[6g_*\Qbar+\sqrt{36(g_*\Qbar)^2-3}\right]^{\frac23}}{3^{\frac23}[6g_*\Qbar+\sqrt{36(g_*\Qbar)^2-3}]^{\frac13}}
\label{mudefa}
\ee
The computation of the leading contribution $\Delta_{-1}$ is entirely analogous to the $U(1)$ case and is given by
\be
\frac{4\Delta_{-1}(g_* \Qbar)}{g_*\Qbar}=\frac{3^{\frac23}[x+\sqrt{x^2-3}]^{\frac13}}{3^{\frac13}+[x+\sqrt{x^2-3}]^{\frac23}}
+\frac{3^{\frac13}\{3^{\frac13}+[x+\sqrt{x^2-3}]^{\frac23}\}}{[x+\sqrt{x^2-3}]^{\frac13}},
\label{DelmQ}
\ee
where $x=6g_*\Qbar$.  Its expansion for small $g_*\Qbar$ takes the form
\be
\frac{\Delta_{-1}(g_* \Qbar)}{g_*}=\Qbar\left[1+\frac13g_*\Qbar-\frac29(g_*\Qbar)^2+\frac{8}{27}(g_*\Qbar)^3
-\frac{14}{27}(g_*\Qbar)^4
+\Ocal\left\{(g_*\Qbar)^5\right\}\right].
\label{lead}
\ee
 The non-leading corrections $\Delta_0$ are once more given by the 
determinant of small fluctuations. There are two modes corresponding to those in the abelian case, with the dispersion relation
\be
\omega_{\pm}^2(l)=J_l^2+3\mu^2-m^2\pm\sqrt{4J_l^2\mu^2+(3\mu^2-m^2)^2}
\label{omdef}
\ee
where
\be
J_l^2=\frac{l(l+d-2)}{R^2}
\label{Jdef}
\ee
is the eigenvalue of the Laplacian on the sphere.
 In addition there are $\frac{N}{2}-1$ ``Type II'' (non-relativistic)\cite{Niel} Goldstone modes and $\frac{N}{2}-1$ massive states with dispersion relation
\be
\omega_{\pm\pm}(l)=\sqrt{J_l^2+\mu^2}\pm\mu,
\label{ompdef}
\ee
with $J_l$ as defined in Eq.~\eqref{Jdef}. 
We then find that $\Delta_0$ is given by
\be
\Delta_0(g_*\Qbar)=\frac{1}{2}\sum_{l=0}^{\infty}\sigma_l
\label{lsum}
\ee
where
\begin{align}
\sigma_l=Rn_l\left\{\omega^*_+(l)+\omega^*_-(l)+\left(\frac{N}{2}-1\right)[\omega^*_{++}(l)+\omega^*_{--}(l)]\right\}.
\label{ONDel}
\end{align}
 Here
\be
n_l=\frac{(2l+d-2)\Gamma(l+d-2)}{\Gamma(l+1)\Gamma(d-1)}
\ee
 is the multiplicity of the laplacian on the $d$-dimensional sphere, and $\omega^*_{\pm}$, $\omega^*_{++}$, $\omega^*_{--}$ are defined as in Eqs.~\eqref{omdef}, \eqref{ompdef} respectively, evaluated at the fixed point with $R$, $\mu_*$ related by Eq.~\eqref{mudefa}.
For the small $(g_*\Qbar)$ computation, we need to isolate the divergent contribution in the sum in Eq.~\eqref{lsum}. We use the large-$l$ expansion of $\sigma_l$,
\be
\sigma_l=\sum_{n=1}^{\infty}c_nl^{d-n}
\label{largel}
\ee
with
\begin{align}
c_1=N,\quad c_2=&3N,\nn
c_3=&\frac{1}{2}[5N-2+(N+2)(R\mu_*)^2],\nn
c_4=&\frac{1}{2}[N-2+(N+2)(R\mu_*)^2],\nn
c_5=&\frac{N+8}{8}(R^2\mu_*^2-1)^2\left[-1+\left(\gamma-\frac32\right)\epsilon\right]\nn
&-\frac{29}{12}(R^2\mu_*^2-1)\epsilon-\left(\frac{11}{24}R^2\mu_*^2-\frac{1}{5}\right)N\epsilon.
\label{cvals}
\end{align}
We can write
\begin{align}
\Delta_0(g_*\Qbar)=&-\frac{15\mu_*^4R^4+6\mu_*^2R^2-5}{16}+\frac12\sum_{l=1}^{\infty}\sigmabar_l+\sqrt{\frac{3\mu_*^2R^2-1}{2}}\nn
&-\frac{1}{16}\left(\frac{N}{2}-1\right)[7-16R\mu_*+6R^2\mu_*^2+3R^4\mu_*^4],
\end{align}
where
\be
\sigmabar_l=\sigma_l-c_1l^3-c_2l^2-c_3l-c_4-c_5\frac1l.
\ee
Here the divergent parts have been isolated and the sums over $l$ performed, as explained in Refs.~\cite{Bad2} and \cite{sann}. The sum over $\frac{1}{l^{d-n}}$ for $n=5$ leads to a pole in $\epsilon$ which cancels against the pole in the bare coupling. The sum over $\sigmabar_l$ is then finite and setting $d=4$ and  expanding in small $g_*\Qbar$ can be performed analytically. We obtain
\begin{align}
\Delta_0=&-\frac16(10+N)g_*\Qbar+\frac{1}{18}(6-N)(g_*\Qbar)^2\nn
&+\frac{1}{27}[N-36+2(14+N)\zeta_3](g_*\Qbar)^3\nn
&-\frac{1}{81}[4(N-73)+2(6N+65)\zeta_3+5(N+30)\zeta_5](g_*\Qbar)^4+\ldots
\label{nonlead}
\end{align}
Adding Eqs.~\eqref{lead} and \eqref{nonlead}, we find \cite{sann}
\begin{align}
\frac{\Delta_{-1}(g_*\Qbar)}{g_*}+\Delta_0(g_*\Qbar)=&
\Qbar+\frac16[2\Qbar-(N+10)]g_*\Qbar-\frac{1}{18}[4\Qbar+(N-6)](g_*\Qbar)^2\nn
&+\frac{1}{27}[8\Qbar+N-36+2(N+14)\zeta_3](g_*\Qbar)^3\nn
&+\Bigl\{-\frac{14}{27}\Qbar-\frac{1}{81}[4(N-73)+2(6N+65)\zeta_3\nn
&+5(N+30)\zeta_5]\Bigr\}(g_*\Qbar)^4+\ldots
\label{Delfull}
\end{align}
\section{The diagrammatic calculation}
In this section we carry out the perturbative calculation to confirm the semiclassical result at leading and next-to-leading order in $\Qbar)$ up to four-loop level, as displayed in Eq.~\eqref{Delfull}.
\begin{figure}[ht]
\center\begin{tikzpicture}
\matrix[column sep = 1cm]
{

\node (vert_cent) {\hspace{-13pt}$\phantom{-}$};
	\vertex at \coord{-90} (A) {};
	\vertex at \coord{90} (B) {};
         \draw  [bend left=70] (A) to (B);
	\draw  [bend left=-70] (A) to (B);
\draw  (B) to (1.5,0.25);
\draw  (B) to (1.5,-0.25);
\vbeta at (A) {};

	;
\draw (0,-1.5) node {\small{(a)}};

&
\node (vert_cent) {\hspace{-13pt}$\phantom{-}$};
	\vertex at \coord{-90} (A) {};
	\vertex at \coord{150} (B) {};
	\vertex at \coord{30} (C) {};
	\vertex at (0.5,0)  (E) {};
         \draw  [bend left=70] (A) to (C);
           \draw  [bend left=-70] (A) to (B) ;
           \draw  [bend left=-10] (A) to (C) ;
	\draw  [bend left=0] (C) to (B);
\draw  (C) to (B);
\draw  (C) to (0.876,1.218); 
\draw  (B) to (0.616,-1.367);
\draw  (B) to (0.876,-1.218); 
\vbeta at (A) {};

	;
\draw (0,-1.5) node {\small{(b)}};

&

\node (vert_cent) {\hspace{-13pt}$\phantom{-}$};
	\vertex at \coord{-90} (A) {};
	\vertex at \coord{150} (B) {};
	\vertex at \coord{30} (C) {};
	\vertex at (0.5,0)  (E) {};
         \draw  [bend left=70] (A) to (C);
           \draw  [bend left=-70] (A) to (B) ;
	\draw  [bend left=30] (C) to (B);
\draw  [bend left=30] (B) to (C);
\draw  (C) to (0.876,1.218); 
\draw  (B) to (0.876,-1.218); 
\vbeta at (A) {};

	;
\draw (0,-1.5) node {\small{(c)}};

\\};
\end{tikzpicture}
\caption{One- and two-loop diagrams for $\gamma_{T_{\Qbar}}$ contributing at leading $n$. Here and elsewhere, the lozenge denotes the $T_{\Qbar}$ vertex.}\label{diagtwo}
\end{figure}
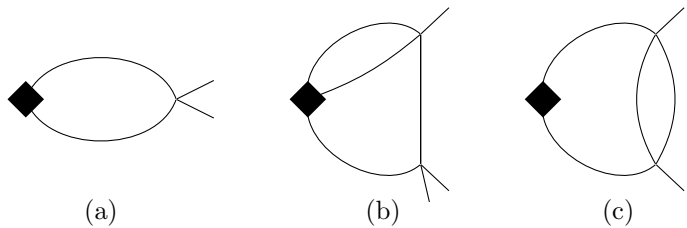

The one-loop contribution to $\gamma_{T_{\Qbar}}$ comes solely from the diagram depicted in Fig.~\ref{diagtwo}(a) and  is given by
\be
\gamma^{(1)}_{T_{\Qbar}}=-\frac13g\Qbar(1-\Qbar).
\label{gamone}
\ee
As mentioned before, the derivation of the semiclassical result relied on working at the conformal fixed point $g_*$. However, surprisingly, at two, three and four loops we will see that the functional forms of the semiclassical and perturbative results agree for general $g$ and not just on substitution  of $g=g_*$ with $g_*$ as given in Eq.~\eqref{gfix}. It is only at one loop where the agreement only holds at the fixed point. Specifically, the leading 
terms $\Qbar\left(\frac d2-1\right)+\gamma^{(1)}_{T_{\Qbar}}$ on the left-hand side of Eq.~\eqref{Tscal} (as given in Eq.~\eqref{gamone}) only agree with the $\Ocal(g^0)$ and $\Ocal(g)$ terms in $\frac{\Delta_{-1}(g\Qbar)}{g}+\Delta_0(g\Qbar)$ on the right-hand side of Eq.~\eqref{Tscal} (as obtained from Eq.~\eqref{Delfull}) after substituting $g=g_*\approx\frac{3\epsilon}{N+8}$. In this case, specialising to the fixed point has induced a mixing between the classical and one-loop $\Ocal(\Qbar)$ terms.

The leading $\Ocal(\Qbar^3)$ two-loop contribution to $\gamma_{T_{\Qbar}}$ comes purely from the diagram depicted in Fig.~\ref{diagtwo}(b) (with three lines emerging from the $T_{\Qbar}$ vertex), while the next-to-leading $\Ocal(\Qbar^2)$ contributions are generated by this diagram together with those in Fig.~\ref{diagtwo}(c) (with two lines emerging from the $T_{\Qbar}$ vertex). The contributions are given by
\begin{align}
\gamma^{(2)}_{(b)}=&-\frac29g^2\Qbar(\Qbar-1)(\Qbar-2),\\
\gamma^{(2)}_{(c)}=&-\frac19g^2\left(3+\frac12N\right)\Qbar(\Qbar-1),
\end{align}
producing leading and next-to-leading terms given by
\be
\gamma^{(2)}_{T_{\Qbar}}=-\frac{1}{18}(g\Qbar)^2(4\Qbar-6+N),
\ee
 in accord with the semiclassical results in Eq.~\eqref{Delfull}. As emphasised earlier, this agreement holds for general $g$ and not just at the conformal fixed point. This is because at two and higher loops, in contrast to what we saw at one loop, specialising to the fixed point $g=g_*$ as given in Eq.~\eqref{gfix} does not induce any mixing between leading or next-to-leading terms at different loop orders. Therefore if Eq.~\eqref{Tscal} holds at the fixed point, it must also hold in general. In fact the agreement was already checked at the fixed point in Ref.~\cite{sann} in the general $O(N)$ case, and in the $U(1)$ case in Ref.~\cite{Bad2}.

\begin{figure}[ht]
\center\begin{tikzpicture}
\matrix[column sep = 1cm]
{

	\node (vert_cent) {\hspace{-13pt}$\phantom{-}$};
	\vertex at \coord{-90} (A) {};
	\vertex at \coord{150} (B) {};
	\vertex at \coord{30} (C) {};
	\vertex at \coord{90} (D) {};
	\draw  [bend left=0] (B) to (D);
	\draw  [bend left=0] (C) to (D);
           \draw  [bend left=-70] (A) to (B) ;
           \draw  [bend left=25] (A) to (B) ;
           \draw  [bend left=-25] (A) to (C) ;
         \draw  [bend left=70] (A) to (C);
\draw  (C) to (0.876,1.218);
\draw  (B) to (0.876,-1.218);
\draw  (D) to (1.35,-0.35);
\draw  (D) to (1.35,0.35);

\vbeta at (A) {};
	;
\draw (0,-1.5) node {\small{(a)}};

&

\node (vert_cent) {\hspace{-13pt}$\phantom{-}$};
	\vertex at \coord{-90} (A) {};
	\vertex at \coord{150} (B) {};
	\vertex at \coord{30} (C) {};
	\vertex at \coord{90} (D) {};
           \draw  (A) to (D);
	\draw  [bend left=0] (D) to (B);
	\draw  [bend left=0] (C) to (D);
           \draw  [bend left=-70] (A) to (B) ;
           \draw  [bend left=-25] (A) to (C) ;
         \draw  [bend left=70] (A) to (C);
\draw  (C) to (0.876,1.218);
\draw  (B) to (0.616,-1.367);
\draw  (B) to (0.876,-1.218); 
\draw  (D) to (1.5,0);

\vbeta at (A) {};
	;
\draw (0,-1.5) node {\small{(b)}};

&

	\node (vert_cent) {\hspace{-13pt}$\phantom{-}$};
	\vertex at \coord{-90} (A) {};
	\vertex at \coord{150} (B) {};
	\vertex at \coord{30} (C) {};
	\vertex at \coord{90} (D) {};
           \draw [bend left=10] (A) to (D);
           \draw [bend left=-10] (A) to (D);
	\draw  [bend left=0] (D) to (B);
	\draw  [bend left=0] (D) to (C);
           \draw  [bend left=-70] (A) to (B) ;
         \draw  [bend left=70] (A) to (C);
\draw  (C) to (0.616,1.367);
\draw  (C) to (0.876,1.218);
\draw  (B) to (0.616,-1.367);
\draw  (B) to (0.876,-1.218);

\vbeta at (A) {};
	;
\draw (0,-1.5) node {\small{(c)}};

\\};
\end{tikzpicture}
\caption{Three-loop diagrams for $\gamma_{T_{\Qbar}}$ contributing at leading $n$}\label{diagthree}
\end{figure}
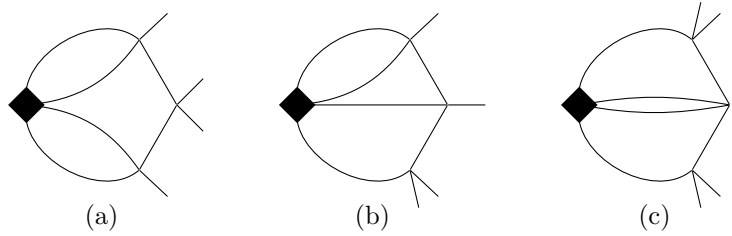

\begin{figure}[ht]
\center\begin{tikzpicture}
\matrix[column sep = 1cm]
{
\node (vert_cent) {\hspace{-13pt}$\phantom{-}$};
	\vertex at \coord{-90} (A) {};
	\vertex at \coord{150} (B) {};
	\vertex at \coord{30} (C) {};
	\vertex at \coord{90} (D) {};
	\draw  [bend left=0] (D) to (B);
	\draw  [bend left=70] (C) to (D);
	\draw  [bend left=-30] (C) to (D);
         \draw  [bend left=70] (A) to (C);
           \draw  [bend left=-75] (A) to (B) ;
           \draw  [bend left=-25] (A) to (C) ;

\draw  (B) to (0.616,-1.367);
\draw  (B) to (0.876,-1.218);
\draw  (D) to (1.5,0);

\vbeta at (A) {};
	;
\draw (0,-1.5) node {\small{(a)}};

&

\node (vert_cent) {\hspace{-13pt}$\phantom{-}$};
	\vertex at \coord{-90} (A) {};
	\vertex at \coord{150} (B) {};
	\vertex at \coord{30} (C) {};
	\vertex at \coord{90} (D) {};
	\draw  [bend left=70] (D) to (B);
	\draw  [bend left=0] (C) to (D);
	\draw  [bend left=30] (B) to (D);
         \draw  [bend left=70] (A) to (C);
           \draw  [bend left=-75] (A) to (B) ;
           \draw  [bend left=-25] (A) to (C) ;
\draw  (C) to (0.876,1.218);
\draw  (D) to (1.5,0);
\draw  (B) to (0.876,-1.218);

\vbeta at (A) {};

	;
\draw (0,-1.5) node {\small{(b)}};

&

	\node (vert_cent) {\hspace{-13pt}$\phantom{-}$};
	\vertex at \coord{-90} (A) {};
	\vertex at \coord{150} (B) {};
	\vertex at \coord{30} (C) {};
	\vertex at \coord{90}  (D) {};
	\vertex at (0.5,0)  (E) {};
         \draw  [bend left=70] (A) to (C);
           \draw  [bend left=-70] (A) to (B) ;
           \draw  [bend left=-10] (A) to (C) ;
	\draw  [bend left=30] (C) to (D);
	\draw  [bend left=-30] (B) to (D);
	\draw  [bend left=0] (C) to (B);
\draw  (B) to (0.876,-1.218);
\draw  (D) to (1.43,-0.25);
\draw  (D) to (1.43,0.25);
\vbeta at (A) {};

	;
\draw (0,-1.5) node {\small{(c)}};

\\

	\node (vert_cent) {\hspace{-13pt}$\phantom{-}$};
	\vertex at \coord{-90} (A) {};
	\vertex at \coord{150} (B) {};
	\vertex at \coord{30} (C) {};
	\vertex at \coord{90} (D) {};
           \draw  (A) to (D);
	\draw  [bend left=0] (D) to (B);
	\draw  [bend left=70] (C) to (D);
	\draw  [bend left=30] (D) to (C);
         \draw  [bend left=70] (A) to (C);
           \draw  [bend left=-70] (A) to (B) ;
\draw  (C) to (0.75,1.3);
\draw  (B) to (0.477,-1.422);
\draw  (B) to (0.993,-1.124);

\vbeta at (A) {};
	;
\draw (0,-1.5) node {\small{(d)}};

&

	\node (vert_cent) {\hspace{-13pt}$\phantom{-}$};
	\vertex at \coord{-90} (A) {};
	\vertex at \coord{150} (B) {};
	\vertex at \coord{30} (C) {};
	\vertex at (0.5,0)  (D) {};
           \draw [bend left=0] (A) to (D);

	\draw  [bend left=0] (B) to (D);
	\draw  [bend left =0] (D) to (C);
         \draw  [bend left=60] (A) to (C);
           \draw  [bend left=-60] (A) to (B) ;

\draw (C) ..controls (1,.86) and (1.2,.2) .. (1.2,0);
\draw (B) ..controls (1,-.86) and (1.2,-.2) .. (1.2,0);
\draw  (C) to (0.616,1.367);
\draw  (B) to (0.616,-1.367);
\draw  (D) to (1,0);

\vbeta at (A) {};

	;
\draw (0,-1.5) node {\small{(e)}};

\\};
\end{tikzpicture}
\caption{Three-loop diagrams for $\gamma_{T_{\Qbar}}$ contributing at next-to-leading $n$}\label{diagthreea}
\end{figure}
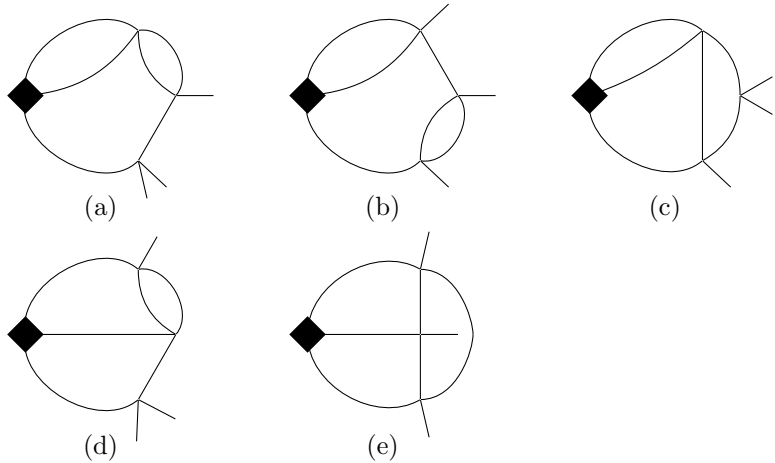

The leading $\Ocal(\Qbar^4)$ three-loop contributions to $\gamma_{T_{\Qbar}}$ come purely from the diagrams depicted in Fig.~\ref{diagthree} (with four lines emerging from the $T_{\Qbar}$ vertex), while the next-to-leading $\Ocal(\Qbar^3)$ contributions are generated by these diagrams together with those in Fig.~\ref{diagthreea} (with three lines emerging from the $T_{\Qbar}$ vertex).

\begin{table}
\begin{center}
\begin{tabular}{|c|c|c|} \hline
&&\\
Graph & Symmetry Factor&Simple Pole\\
&&\\
\hline
&&\\
2(a)&$\frac{1}{54}\frac{\Qbar!}{(\Qbar-4)!}$&$-\frac23$\\
&&\\
\hline
&&\\
2(b)&$\frac{2}{27}\frac{\Qbar!}{(\Qbar-4)!}$&$\frac43$\\
&&\\
\hline
&&\\
2(c)&$\frac{1}{54}\frac{\Qbar!}{(\Qbar-4)!}$&$\frac23$\\
&&\\
\hline
&&\\
3(a)&$\frac{1}{27}\frac{\Qbar!}{(\Qbar-3)!}$&$-\frac23$\\
&&\\
\hline
&&\\
3(b)&$\frac{4}{27}\frac{\Qbar!}{(\Qbar-3)!}A$&$-\frac23$\\
&&\\
\hline
&&\\
3(c)&$\frac{2}{27}\frac{\Qbar!}{(\Qbar-3)!}$&$\frac43$\\
&&\\
\hline
&&\\
3(d)&$\frac{4}{27}\frac{\Qbar!}{(\Qbar-3)!}A$&$\frac43$\\
&&\\
\hline
&&\\
3(e)&$\frac{8}{27}\frac{\Qbar!}{(\Qbar-3)!}B$&$4\zeta_3$\\
&&\\
\hline
\end{tabular}
\caption{\label{3loop}Three-loop results from Figs.~\ref{diagthree}, \ref{diagthreea}}
\end{center}
\end{table}

The simple pole contributions from individual three loop diagrams may be extracted from Ref.~\cite{kaz} and are listed in Table~\ref{3loop}, together with the corresponding symmetry factor. A factor of $g^3$ is understood in each case. The $N$-dependent factors $A$ and $B$ are given by
\be
A=\frac18(N+6),\quad B=\frac{1}{16}(N+14).
\ee
When added and multiplied by a loop factor of 3, the leading and non-leading three-loop contributions to  $\gamma_{T_{\Qbar}}$ are found to be 
\be
\gamma^{(3)}_{T_{\Qbar}}=\frac{1}{27}(g\Qbar)^3[8\Qbar+N-36+2(14+N)\zeta_3],
\ee
once again in accord with the semiclassical results in Eqs.~\eqref{Delfull}, for general $g$. Equivalently, this agreement was already checked at the fixed point in Ref.~\cite{sann}.

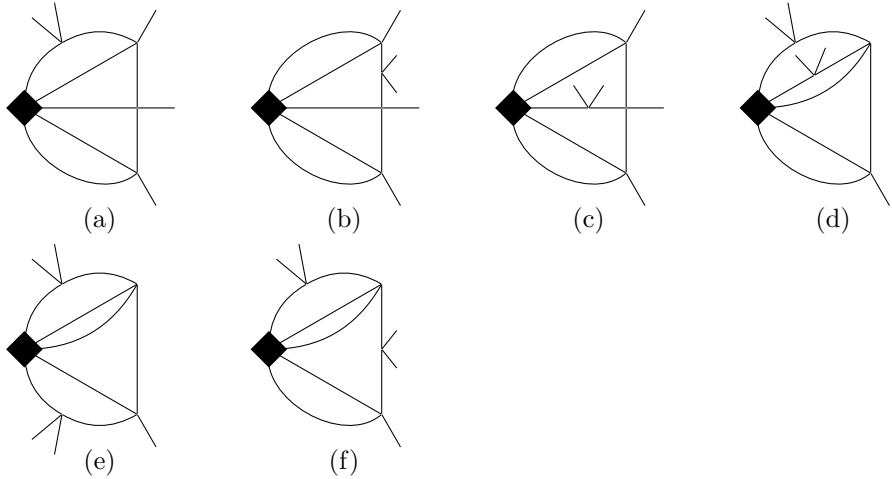
\begin{figure}[ht]
\center\begin{tikzpicture}
\matrix[column sep = 1cm]
{

	%\node (vert_cent) {\hspace{-13pt}$\phantom{-}$};
	\vertex at \coord{-90} (A) {};
	\vertex at \coord{150} (B) {};
	\vertex at \coord{30} (C) {};
	\vertex at (0.5,0) (D) {};
	\vertex at \coord{-30} (E) {};
           \draw [bend left=0] (A) to (D);
	\draw  [bend left=0] (B) to (D);
	\draw  [bend left=0] (D) to (C);
           \draw  [bend left=-70] (A) to (B) ;
           \draw  [bend left=0] (A) to (B) ;
           \draw  [bend left=0] (A) to (C) ;
	\draw  [bend left=30] (A) to (E);
	\draw  [bend left=-30] (C) to (E);
\draw  (D) to (1,0);
 \draw  (B) to (0.75,-1.3);
 \draw  (C) to (0.75,1.3);
\vbeta at (A) {};
\draw  (E) to (-.9,1.2);
\draw  (E) to (-.6,1.4);
\draw (0,-1.5) node {\small{(a)}};
	
&

	\node (vert_cent) {\hspace{-13pt}$\phantom{-}$};
	\vertex at \coord{-90} (A) {};
	\vertex at \coord{150} (B) {};
	\vertex at \coord{30} (C) {};
	\vertex at  (0.5,0) (D) {};
	\vertex at  (0.5,0.466) (E) {};
           \draw [bend left=0] (A) to (D);
	\draw  [bend left=0] (B) to (D);
           \draw  [bend left=-70] (A) to (B) ;
           \draw  [bend left=0] (A) to (B) ;
           \draw  [bend left=0] (A) to (C) ;
           \draw  [bend left=70] (A) to (C) ;
	\draw   (D) to (E);
	\draw   (C) to (E);
\draw  (D) to (1,0);
 \draw  (B) to (0.75,-1.3);
 \draw  (C) to (0.75,1.3);
\vbeta at (A) {};
\draw  (E) to (.7,.7);
\draw  (E) to (.7,.2);
\draw (0,-1.5) node {\small{(b)}};

&

	\node (vert_cent) {\hspace{-13pt}$\phantom{-}$};
	\vertex at \coord{-90} (A) {};
	\vertex at \coord{150} (B) {};
	\vertex at \coord{30} (C) {};
	\vertex at  (0.5,0) (D) {};
	\vertex at  (0,0) (E) {};
	\draw  [bend left=0] (B) to (D);
           \draw  [bend left=-70] (A) to (B) ;
           \draw  [bend left=0] (A) to (B) ;
           \draw  [bend left=0] (A) to (C) ;
           \draw  [bend left=70] (A) to (C) ;
	\draw   (D) to (E);
	\draw   (A) to (E);
	\draw   (C) to (D);
\draw  (D) to (1,0);
 \draw  (B) to (0.75,-1.3);
 \draw  (C) to (0.75,1.3);
\vbeta at (A) {};
\draw  (E) to (-0.2,.3);
\draw  (E) to (0.2,.3);
\draw (0,-1.5) node {\small{(c)}};

&

	%\node (vert_cent) {\hspace{-13pt}$\phantom{-}$};
	\vertex at \coord{-90} (A) {};
	\vertex at \coord{150} (B) {};
	\vertex at \coord{30} (C) {};
	\vertex at (0.5,0) (D) {};
	\vertex at \coord{-30} (E) {};
	\vertex at (-0.25,0.433) (F) {};
           \draw [bend left=-30] (A) to (C);
	\draw  [bend left=0] (B) to (C);
           \draw  [bend left=-70] (A) to (B) ;
           \draw  [bend left=0] (A) to (B) ;
	\draw  [bend left=30] (A) to (E);
	\draw  [bend left=-30] (C) to (E);
	\draw   (A) to (F);
	\draw   (C) to (F);
 \draw  (B) to (0.75,-1.3);
\vbeta at (A) {};
\draw  (E) to (-.9,1.2);
\draw  (E) to (-.6,1.4);
\draw  (F) to (-0.1,0.8);
\draw  (F) to (-0.5,0.7);

\draw (0,-1.5) node {\small{(d)}};

\\

	%\node (vert_cent) {\hspace{-13pt}$\phantom{-}$};
	\vertex at \coord{-90} (A) {};
	\vertex at \coord{150} (B) {};
	\vertex at \coord{30} (C) {};
	\vertex at (0.5,0) (D) {};
	\vertex at \coord{-30} (E) {};
	\vertex at \coord{-150} (F) {};
	\draw  [bend left=30] (A) to (E);
	\draw  [bend left=-30] (C) to (E);
           \draw  (A) to (C);
           \draw [bend left=-30] (A) to (C);
	\draw  [bend left=0] (C) to (B);
           \draw  [bend left=0] (A) to (B) ;
	\draw  [bend left=-30] (A) to (F);
	\draw  [bend left=30] (B) to (F);
 \draw  (B) to (0.75,-1.3);
\vbeta at (A) {};
\draw  (E) to (-.9,1.2);
\draw  (E) to (-.6,1.4);
\draw  (F) to (-.9,-1.2);
\draw  (F) to (-.6,-1.4);

\draw (0,-1.5) node {\small{(e)}};

&

%\node (vert_cent) {\hspace{-13pt}$\phantom{-}$};
	\vertex at \coord{-90} (A) {};
	\vertex at \coord{150} (B) {};
	\vertex at \coord{30} (C) {};
	\vertex at (0.5,0) (D) {};
	\vertex at \coord{-30} (E) {};
	\vertex at \coord{-150} (F) {};
	\draw  [bend left=30] (A) to (E);
	\draw  [bend left=-30] (C) to (E);
           \draw  (A) to (C);
           \draw [bend left=-30] (A) to (C);
	\draw  [bend left=0] (C) to (D);
	\draw  [bend left=0] (B) to (D);
           \draw  [bend left=0] (A) to (B) ;
	\draw  [bend left=-70] (A) to (B);
 \draw  (B) to (0.75,-1.3);
\vbeta at (A) {};
\draw  (E) to (-.9,1.2);
\draw  (E) to (-.6,1.4);
\draw  (D) to (0.7,0.25);
\draw  (D) to (0.7,-0.25);

\draw (0,-1.5) node {\small{(f)}};

\\};
\end{tikzpicture}
\caption{Four-loop diagrams for $\gamma_{T_{\Qbar}}$ contributing at leading $n$}\label{diagfour}
\end{figure}

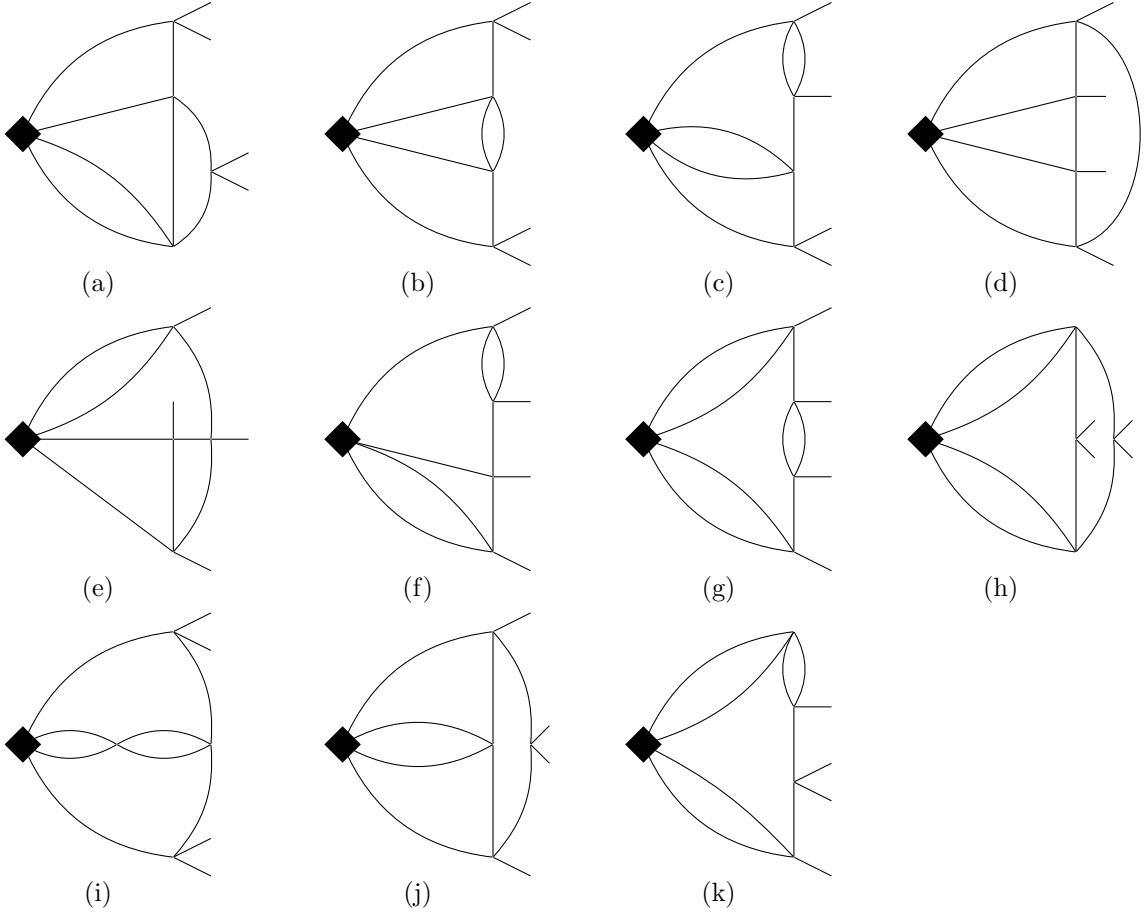
\begin{figure}[ht]
\center\begin{tikzpicture}
\matrix[column sep = 1cm]
{

\vertex at (-1,0) (A) {};
\vertex at (1,1.5) (B) {};
\vertex at (1,0.5) (C) {};
\vertex at (1.5,-0.5) (D) {};
\vertex at (1,-1.5) (E) {};

	\draw  [bend left=30] (A) to (B);
	\draw   (C) to (B);
	\draw   (E) to (C);
	\draw  [bend left=30] (C) to (D);
	\draw  [ bend left=-30] (E) to (D);
\draw   (D) to (2,-0.25);
\draw   (D) to (2,-0.75);
\draw   (B) to (1.5,1.75);
	\draw  [bend left=-30] (A) to (E);
	\draw  [bend left=20] (A) to (E);
	\draw   (A) to (C);
\draw   (B) to (1.5,1.25);
\vbeta at (A) {};

\draw (0,-2) node {\small{(a)}};

&

\vertex at (-1,0) (A) {};
\vertex at (1,1.5) (B) {};
\vertex at (1,0.5) (C) {};
\vertex at (1,-0.5) (D) {};
\vertex at (1,-1.5) (E) {};

	\draw  [bend left=30] (A) to (B);
	\draw   (A) to (D);
	\draw  [bend left=30] (C) to (D);
	\draw  [bend left=30] (D) to (C);
	\draw   (C) to (B);
	\draw   (D) to (E);
\draw   (B) to (1.5,1.25);
\draw   (B) to (1.5,1.75);
\draw   (E) to (1.5,-1.75);
\draw   (E) to (1.5,-1.25);
	\draw   (A) to (C);
	\draw  [bend left=-30] (A) to (E);
\vbeta at (A) {};

\draw (0,-2) node {\small{(b)}};

&

\vertex at (-1,0) (A) {};
\vertex at (1,1.5) (B) {};
\vertex at (1,0.5) (C) {};
\vertex at (1,-0.5) (D) {};
\vertex at (1,-1.5) (E) {};

	\draw  [bend left=30] (A) to (B);
	\draw  [bend left=30] (A) to (D);
	\draw  [bend left=30] (B) to (C);
	\draw  [bend left=30] (C) to (B);
	\draw   (D) to (C);
	\draw   (D) to (E);
\draw   (C) to (1.5,0.5);
\draw   (B) to (1.5,1.75);
\draw   (E) to (1.5,-1.75);
\draw   (E) to (1.5,-1.25);
	\draw  [bend left=-30] (A) to (D);
	\draw  [bend left=-30] (A) to (E);
\vbeta at (A) {};

\draw (0,-2) node {\small{(c)}};

&

\vertex at (-1,0) (A) {};
\vertex at (1,1.5) (B) {};
\vertex at (1,0.5) (C) {};
\vertex at (1.5,0.5) (F) {};
\vertex at (1,-0.5) (D) {};
\vertex at (1,-1.5) (E) {};

	\draw  [bend left=30] (A) to (B);
	\draw   (A) to (C);
	\draw   (A) to (D);
	\draw  [bend left=-75] (E) to (B);
	\draw   (B) to (C);
	\draw   (C) to (D);
	\draw   (D) to (E);
\draw   (C) to (1.4,0.5);
\draw   (D) to (1.4,-0.5);
\draw   (B) to (1.5,1.75);
\draw   (E) to (1.5,-1.75);

	\draw  [bend left=-30] (A) to (E);
\vbeta at (A) {};

\draw (0,-2) node {\small{(d)}};

\\

\vertex at (-1,0) (A) {};
\vertex at (1,1.5) (B) {};
\vertex at (1,0) (C) {};
\vertex at (1,-1.5) (D) {};
\vertex at (1.5,0) (E) {};

	\draw  [bend left=30] (A) to (B);
	\draw  [bend left=-20] (A) to (B);
	\draw  [bend left=23] (B) to (E);
	\draw  [bend left=-23] (D) to (E);
	\draw   (C) to (D);
	\draw   (A) to (C);
	\draw   (E) to (C);
	\draw   (A) to (D);
\draw   (B) to (1.5,1.75);
\draw   (D) to (1.5,-1.75);
\draw   (E) to (2,0);
\draw   (C) to (1,0.5);
\vbeta at (A) {};

\draw (0,-2) node {\small{(e)}};

&

\vertex at (-1,0) (A) {};
\vertex at (1,1.5) (B) {};
\vertex at (1,0.5) (C) {};
\vertex at (1,-0.5) (D) {};
\vertex at (1,-1.5) (E) {};

	\draw  [bend left=30] (A) to (B);
	\draw  [bend left=20] (A) to (E);
	\draw  [bend left=30] (B) to (C);
	\draw  [bend left=30] (C) to (B);
	\draw   (D) to (C);
	\draw   (E) to (D);
\draw   (C) to (1.5,0.5);
\draw   (D) to (1.5,-0.5);
\draw   (E) to (1.5,-1.75);
\draw   (B) to (1.5,1.75);
	\draw  [bend left=-30] (A) to (E);
	\draw   (A) to (D);
\vbeta at (A) {};

\draw (0,-2) node {\small{(f)}};

&

\vertex at (-1,0) (A) {};
\vertex at (1,1.5) (B) {};
\vertex at (1,0.5) (C) {};
\vertex at (1,-0.5) (D) {};
\vertex at (1,-1.5) (E) {};

	\draw  [bend left=30] (A) to (B);
	\draw  [bend left=-20] (A) to (B);
	\draw  [bend left=30] (C) to (D);
	\draw  [bend left=30] (D) to (C);
	\draw   (B) to (C);
	\draw   (E) to (D);
\draw   (C) to (1.5,0.5);
\draw   (B) to (1.5,1.75);
\draw   (E) to (1.5,-1.75);
\draw   (D) to (1.5,-0.5);
	\draw  [bend left=20] (A) to (E);
	\draw  [bend left=-30] (A) to (E);
\vbeta at (A) {};

\draw (0,-2) node {\small{(g)}};

&

\vertex at (-1,0) (A) {};
\vertex at (1,1.5) (B) {};
\vertex at (1,0) (C) {};
\vertex at (1,-1.5) (D) {};
\vertex at (1.5,0) (E) {};

	\draw  [bend left=30] (A) to (B);
	\draw  [bend left=-20] (A) to (B);
	\draw  [bend left=23] (B) to (E);
	\draw  [bend left=-23] (D) to (E);
	\draw   (D) to (C);
	\draw   (B) to (C);
	\draw  [bend left=-30] (A) to (D);
	\draw  [bend left=20] (A) to (D);
\draw   (E) to (1.75,0.25);
\draw   (E) to (1.75,0.-0.25);
\draw   (C) to (1.25,0.25);
\draw   (C) to (1.25,0.-0.25);
\vbeta at (A) {};

\draw (0,-2) node {\small{(h)}};

\\

\vertex at (-1,0) (A) {};
\vertex at (1,1.5) (B) {};
\vertex at (0.25,0) (C) {};
\vertex at (1,-1.5) (D) {};
\vertex at (1.5,0) (E) {};

	\draw  [bend left=30] (A) to (B);
	\draw  [bend left=-23] (E) to (B);
	\draw  [bend left=23] (E) to (D);
	\draw  [bend left=30] (A) to (C);
	\draw  [bend left=-30] (A) to (C);
	\draw  [bend left=30] (C) to (E);
	\draw  [bend left=-30] (C) to (E);
	\draw  [bend left=-30] (A) to (D);
\draw   (B) to (1.5,1.75);
\draw   (B) to (1.5,1.25);
\draw   (D) to (1.5,-1.75);
\draw   (D) to (1.5,-1.25);
\vbeta at (A) {};

\draw (0,-2) node {\small{(i)}};

&

\vertex at (-1,0) (A) {};
\vertex at (1,1.5) (B) {};
\vertex at (1,0) (C) {};
\vertex at (1,-1.5) (D) {};
\vertex at (1.5,0) (E) {};

	\draw  [bend left=30] (A) to (B);
	\draw  [bend left=30] (A) to (C);
	\draw  [bend left=23] (B) to (E);
	\draw  [bend left=-23] (D) to (E);
	\draw   (C) to (D);
	\draw   (C) to (B);
	\draw  [bend left=-30] (A) to (D);
	\draw  [bend left=-30] (A) to (C);
\draw   (E) to (1.75,0.25);
\draw   (E) to (1.75,0.-0.25);
\draw   (B) to (1.5,1.75);
\draw   (D) to (1.5,-1.75);
\vbeta at (A) {};

\draw (0,-2) node {\small{(j)}};

&

\vertex at (-1,0) (A) {};
\vertex at (1,1.5) (B) {};
\vertex at (1,0.5) (C) {};
\vertex at (1,-0.5) (D) {};
\vertex at (1,-1.5) (E) {};

	\draw  [bend left=30] (A) to (B);
	\draw  [bend left=-20] (A) to (B);
	\draw  [bend left=30] (B) to (C);
	\draw  [bend left=-30] (B) to (C);
	\draw   (C) to (D);
	\draw   (E) to (D);
\draw   (C) to (1.5,0.5);
\draw   (D) to (1.5,-0.25);
\draw   (D) to (1.5,-0.75);
\draw   (E) to (1.5,-1.75);
	\draw  [bend left=-30] (A) to (E);
	\draw  [bend left=10] (A) to (E);
\vbeta at (A) {};

\draw (0,-2) node {\small{(k)}};

\\};
\end{tikzpicture}
\caption{Four-loop diagrams for $\gamma_{T_{\Qbar}}$ contributing at next-to-leading $n$}\label{diagfoura}
\end{figure}

\begin{figure}[ht]
\center\begin{tikzpicture}
\matrix[column sep = 1cm]
{

\vertex at (-1,0) (A) {};
\vertex at (1,1.5) (B) {};
\vertex at (1,0.5) (C) {};
\vertex at (1,-0.5) (D) {};
\vertex at (1,-1.5) (E) {};

	\draw  [bend left=30] (A) to (B);
	\draw  [bend left=-20] (A) to (B);
	\draw  [bend left=30] (C) to (D);
	\draw  [bend left=30] (D) to (C);
	\draw   (B) to (C);
	\draw   (D) to (E);
\draw   (C) to (1.5,0.5);
\draw   (B) to (1.5,1.75);
\draw   (E) to (1.5,-1.75);
\draw   (E) to (1.5,-1.25);
	\draw   (A) to (D);
	\draw  [bend left=-30] (A) to (E);
\vbeta at (A) {};

\draw (0,-2) node {\small{(a)}};

&

\vertex at (-1,0) (A) {};
\vertex at (1,1.5) (B) {};
\vertex at (1,0.5) (C) {};
\vertex at (1.5,0.5) (F) {};
\vertex at (1,-0.5) (D) {};
\vertex at (1,-1.5) (E) {};

	\draw  [bend left=30] (A) to (B);
	\draw   (A) to (C);
	\draw  [bend left=-75] (D) to (B);
	\draw   (B) to (C);
	\draw   (C) to (D);
	\draw   (D) to (E);
\draw   (C) to (1.4,0.5);
\draw   (B) to (1.5,1.75);
\draw   (E) to (1.5,-1.75);
\draw   (E) to (1.5,-1.25);
	\draw   (A) to (D);
	\draw  [bend left=-30] (A) to (E);
\vbeta at (A) {};

\draw (0,-2) node {\small{(b)}};

&

\vertex at (-1,0) (A) {};
\vertex at (1,1.5) (B) {};
\vertex at (1,0) (C) {};
\vertex at (1,-1.5) (D) {};
\vertex at (0,0) (E) {};

	\draw  [bend left=30] (A) to (B);
	\draw  [bend left=-20] (A) to (B);
	\draw  [bend left=30] (D) to (C);
	\draw  [bend left=30] (C) to (D);
	\draw   (B) to (C);
	\draw   (A) to (E);
	\draw   (A) to (D);
	\draw   (C) to (E);
\draw   (B) to (1.5,1.75);
\draw   (D) to (1.5,-1.75);
\draw   (E) to (-0.25,-0.25);
\draw   (E) to (0.25,0.-0.25);
\vbeta at (A) {};

\draw (0,-2) node {\small{(c)}};

&

\vertex at (-1,0) (A) {};
\vertex at (1,1.5) (B) {};
\vertex at (1,0.5) (C) {};
\vertex at (1,-0.5) (D) {};
\vertex at (1,-1.5) (E) {};

	\draw  [bend left=30] (A) to (B);
	\draw  [bend left=-20] (A) to (B);
	\draw  [bend left=30] (C) to (D);
	\draw  [bend left=-30] (C) to (D);
	\draw   (B) to (C);
	\draw   (D) to (E);
\draw   (D) to (1.5,-0.5);
\draw   (B) to (1.5,1.75);
\draw   (E) to (1.5,-1.75);
\draw   (E) to (1.5,-1.25);
	\draw   (A) to (C);
	\draw  [bend left=-30] (A) to (E);
\vbeta at (A) {};

\draw (0,-2) node {\small{(d)}};

\\

\vertex at (-1,0) (A) {};
\vertex at (1,1.5) (B) {};
\vertex at (1,0.5) (C) {};
\vertex at (1,-0.5) (D) {};
\vertex at (1,-1.5) (E) {};

	\draw  [bend left=30] (A) to (B);
	\draw  [bend left=-20] (A) to (B);
	\draw  [bend left=30] (D) to (E);
	\draw  [bend left=30] (E) to (D);
	\draw   (B) to (C);
	\draw   (D) to (C);
	\draw   (A) to (D);
\draw   (C) to (1.5,0.25);
\draw   (C) to (1.5,0.75);
	\draw  [bend left=-30] (A) to (E);
\draw   (B) to (1.5,1.75);
\draw   (E) to (1.5,-1.75);

\vbeta at (A) {};

\draw (0,-2) node {\small{(e)}};

&

\vertex at (-1,0) (A) {};
\vertex at (1,1.5) (B) {};
\vertex at (1,0.5) (C) {};
\vertex at (1,-0.5) (D) {};
\vertex at (1,-1.5) (E) {};

	\draw  [bend left=30] (A) to (B);
	\draw  [bend left=-20] (A) to (B);
	\draw  [bend left=30] (B) to (C);
	\draw  [bend left=-30] (B) to (C);
	\draw   (C) to (D);
	\draw   (D) to (E);
\draw   (C) to (1.5,0.5);
\draw   (D) to (1.5,-0.5);
\draw   (E) to (1.5,-1.75);
\draw   (E) to (1.5,-1.25);
	\draw  [bend left=-30] (A) to (E);
	\draw   (A) to (D);
\vbeta at (A) {};

\draw (0,-2) node {\small{(f)}};

&

\vertex at (-1,0) (A) {};
\vertex at (1,1.5) (B) {};
\vertex at (1,0.5) (C) {};
\vertex at (1.5,-0.5) (D) {};
\vertex at (1,-1.5) (E) {};

	\draw  [bend left=30] (A) to (B);
	\draw  [bend left=-20] (A) to (B);
	\draw   (B) to (C);
	\draw   (E) to (C);
	\draw  [bend left=30] (C) to (D);
	\draw  [ bend left=-30] (E) to (D);
\draw   (C) to (1.5,0.5);
\draw   (D) to (2,-0.25);
\draw   (D) to (2,-0.75);
\draw   (B) to (1.5,1.75);
	\draw  [bend left=-30] (A) to (E);
	\draw  [bend left=20] (A) to (E);
\vbeta at (A) {};

\draw (0,-2) node {\small{(g)}};

&

\vertex at (-1,0) (A) {};
\vertex at (1,1.5) (B) {};
\vertex at (1,0.5) (C) {};
\vertex at (1.5,0.5) (F) {};
\vertex at (1,-0.5) (D) {};
\vertex at (1,-1.5) (E) {};

	\draw  [bend left=30] (A) to (B);
	\draw  [bend left=30] (A) to (C);
	\draw  [bend left=-30] (A) to (C);
	\draw  [bend left=75] (B) to (D);
	\draw   (C) to (B);
	\draw   (C) to (D);
	\draw   (D) to (E);
\draw   (D) to (1.4,-0.5);
\draw   (B) to (1.5,1.75);
\draw   (E) to (1.5,-1.75);
\draw   (E) to (1.5,-1.25);

	\draw  [bend left=-30] (A) to (E);
\vbeta at (A) {};

\draw (0,-2) node {\small{(h)}};

\\

\vertex at (-1,0) (A) {};
\vertex at (1,1.5) (B) {};
\vertex at (1,0) (C) {};
\vertex at (1,-1.5) (D) {};
\vertex at (1.5,0) (E) {};

	\draw  [bend left=30] (A) to (B);
	\draw  [bend left=-20] (A) to (B);
	\draw  [bend left=23] (B) to (E);
	\draw  [bend left=-23] (D) to (E);
	\draw   (C) to (D);
	\draw   (B) to (C);
	\draw   (A) to (C);
	\draw   (A) to (D);
\draw   (C) to (1.3,0);
\draw   (D) to (1.5,-1.75);
\draw   (E) to (1.75,0.25);
\draw   (E) to (1.75,0.-0.25);
\vbeta at (A) {};

\draw (0,-2) node {\small{(i)}};

&

\vertex at (-1,0) (A) {};
\vertex at (1,1.5) (B) {};
\vertex at (1,0.5) (C) {};
\vertex at (1,-0.5) (D) {};
\vertex at (1,-1.5) (E) {};

	\draw  [bend left=30] (A) to (B);
	\draw   (A) to (D);
	\draw  [bend left=30] (B) to (C);
	\draw  [bend left=30] (C) to (B);
	\draw   (C) to (D);
	\draw   (D) to (E);
\draw   (D) to (1.5,-0.5);
\draw   (B) to (1.5,1.75);
\draw   (E) to (1.5,-1.75);
\draw   (E) to (1.5,-1.25);
	\draw   (A) to (C);
	\draw  [bend left=-30] (A) to (E);
\vbeta at (A) {};

\draw (0,-2) node {\small{(j)}};

&

\vertex at (-1,0) (A) {};
\vertex at (1,1.5) (B) {};
\vertex at (1,0.5) (C) {};
\vertex at (1.5,-0.5) (D) {};
\vertex at (1,-1.5) (E) {};

	\draw  [bend left=30] (A) to (B);
	\draw   (B) to (C);
	\draw   (C) to (E);
	\draw  [bend left=30] (C) to (D);
	\draw  [ bend left=-30] (E) to (D);
\draw   (D) to (2,-0.25);
\draw   (D) to (2,-0.75);
\draw   (B) to (1.5,1.75);
	\draw  [bend left=-30] (A) to (E);
	\draw  [bend left=-20] (A) to (B);
	\draw   (A) to (C);
\draw   (E) to (1.5,-1.75);
\vbeta at (A) {};

\draw (0,-2) node {\small{(k)}};

\\};
\end{tikzpicture}
\caption{Four-loop diagrams for $\gamma_{T_{\Qbar}}$ contributing at next-to-leading $n$ (continued)}\label{diagfourb}
\end{figure}
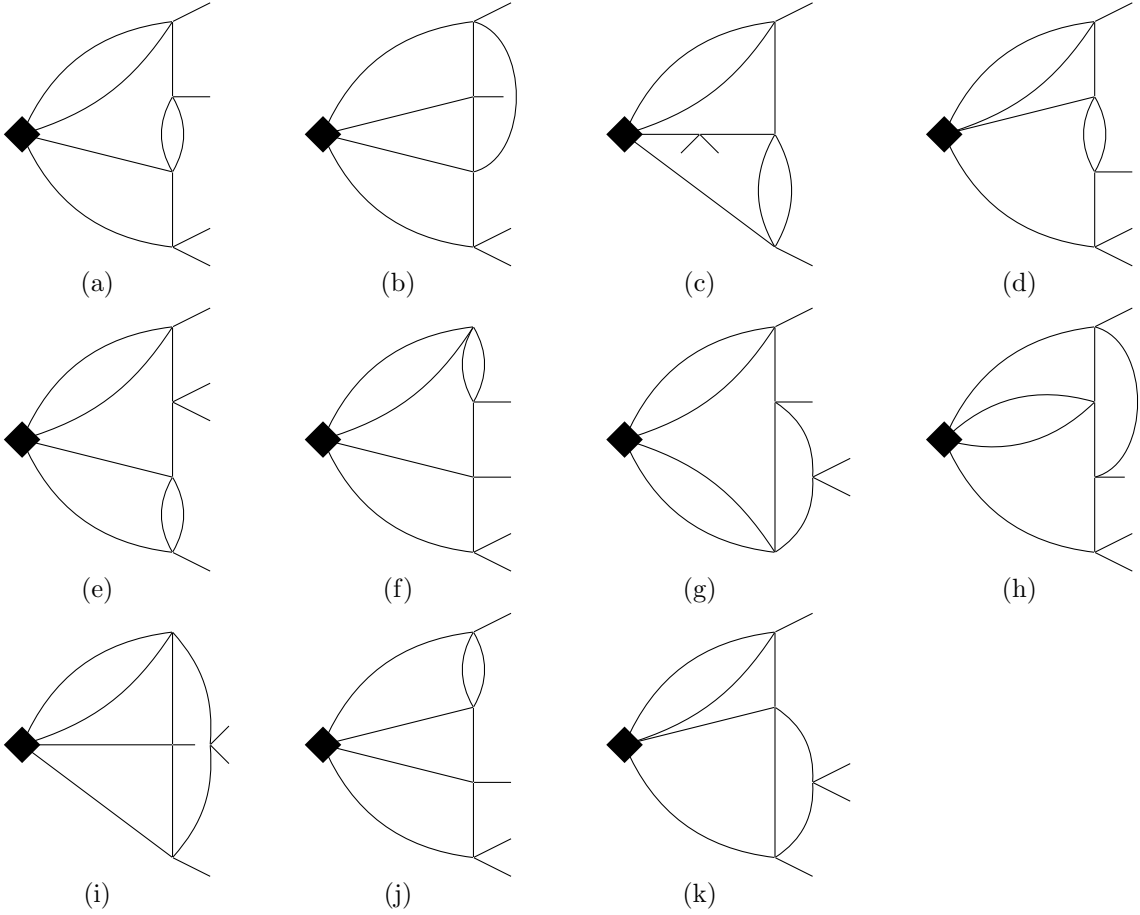

The leading $\Ocal(\Qbar^5)$ four-loop contributions to $\gamma_{T_{\Qbar}}$ come purely from the diagrams depicted in Fig.~\ref{diagfour} (with five lines emerging from the $T_{\Qbar}$ vertex), while the next-to-leading $\Ocal(\Qbar^4)$ contributions are generated by these diagrams together with those in Figs.~\ref{diagfoura} and \ref{diagfourb} (with four lines emerging from the $T_{\Qbar}$ vertex).

\begin{table}
\begin{center}
\begin{tabular}{|c|c|c|} \hline
&&\\
Graph & Symmetry Factor&Simple Pole\\
&&\\
\hline
&&\\
4(a)&$\frac{4}{81}\frac{\Qbar!}{(\Qbar-5)!}$&$\frac52$\\
&&\\
\hline
&&\\
4(b)&$\frac{2}{81}\frac{\Qbar!}{(\Qbar-5)!}$&$-\frac23$\\
&&\\
\hline
&&\\
4(c)&$\frac{1}{81}\frac{\Qbar!}{(\Qbar-5)!}$&$-\frac56$\\
&&\\
\hline
&&\\
4(d)&$\frac{1}{81}\frac{\Qbar!}{(\Qbar-5)!}$&$\frac{11}{6}$\\
&&\\
\hline
&&\\
4(e)&$\frac{2}{81}\frac{\Qbar!}{(\Qbar-5)!}$&$\frac23$\\
&&\\
\hline
&&\\
4(f)&$\frac{1}{81}\frac{\Qbar!}{(\Qbar-5)!}$&$-\frac12$\\
&&\\
\hline
\end{tabular}
\caption{\label{4loop}Four-loop results from Fig.~\ref{diagfour}}
\end{center}
\end{table}

\begin{table}
\begin{center}
\begin{tabular}{|c|c|c|} \hline
&&\\
Graph & Symmetry Factor&Simple Pole\\
&&\\
\hline
&&\\
5(a)&$\frac{2}{81}\frac{\Qbar!}{(\Qbar-4)!}$&$\frac{1}{6}(11-6\zeta_3)$\\
&&\\
\hline
&&\\
5(b)&$\frac{2}{81}\frac{\Qbar!}{(\Qbar-4)!}A$&$\frac{1}{6}(11-6\zeta_3)$\\
&&\\
\hline
&&\\
5(c)&$\frac{4}{81}\frac{\Qbar!}{(\Qbar-4)!}A$&$-\frac{1}{2}$\\
&&\\
\hline
&&\\
5(d)&$\frac{16}{81}\frac{\Qbar!}{(\Qbar-4)!}C$&$10\zeta_5$\\
&&\\
\hline
&&\\
5(e)&$\frac{8}{81}\frac{\Qbar!}{(\Qbar-4)!}B$&$\frac32(2\zeta_3-\zeta_4)$\\
&&\\
\hline
&&\\
5(f)&$\frac{8}{81}\frac{\Qbar!}{(\Qbar-4)!}A$&$-\frac23$\\
&&\\
\hline
&&\\
5(g)&$\frac{2}{81}\frac{\Qbar!}{(\Qbar-4)!}A$&$\frac12(1-2\zeta_3)$\\
&&\\
\hline
&&\\
5(h)&$\frac{1}{324}\frac{\Qbar!}{(\Qbar-4)!}$&$-2(1-\zeta_3)$\\&&\\
\hline
&&\\
5(i)&$\frac{1}{162}\frac{\Qbar!}{(\Qbar-4)!}$&$-2(1-\zeta_3)$\\
&&\\
\hline
&&\\
5(j)&$\frac{2}{81}\frac{\Qbar!}{(\Qbar-4)!}$&$-(1-2\zeta_3)$\\
&&\\
\hline
&&\\
5(k)&$\frac{1}{81}\frac{\Qbar!}{(\Qbar-4)!}$&$\frac12(1-2\zeta_3)$\\
&&\\
\hline
\end{tabular}
\caption{\label{4loopa}Four-loop results from Fig.~\ref{diagfoura}}
\end{center}
\end{table}

\begin{table}
\begin{center}
\begin{tabular}{|c|c|c|} \hline
&&\\
Graph & Symmetry Factor&Simple Pole\\
&&\\
\hline
&&\\
6(a)&$\frac{4}{81}\frac{\Qbar!}{(\Qbar-4)!}A$&$-\frac{1}{6}(5-6\zeta_3)$\\
&&\\
\hline
&&\\
6(b)&$\frac{8}{81}\frac{\Qbar!}{(\Qbar-4)!}B$&$\frac32(2\zeta_3+\zeta_4)$\\
&&\\
\hline
&&\\
6(c)&$\frac{4}{81}\frac{\Qbar!}{(\Qbar-4)!}A$&$-\frac{5}{6}$\\
&&\\
\hline
&&\\
6(d)&$\frac{2}{81}\frac{\Qbar!}{(\Qbar-4)!}$&$-\frac23$\\
&&\\
\hline
&&\\
6(e)&$\frac{4}{81}\frac{\Qbar!}{(\Qbar-4)!}A$&$-\frac23$\\
&&\\
\hline
&&\\
6(f)&$\frac{2}{81}\frac{\Qbar!}{(\Qbar-4)!}$&$-\frac{1}{6}(5-6\zeta_3)$\\
&&\\
\hline
&&\\
6(g)&$\frac{2}{81}\frac{\Qbar!}{(\Qbar-4)!}$&$-\frac{1}{6}(5-6\zeta_3)$\\
&&\\
\hline
&&\\
6(h)&$\frac{4}{81}\frac{\Qbar!}{(\Qbar-4)!}$&$\frac{1}{2}(5-4\zeta_3)$\\
&&\\
\hline
&&\\
6(i)&$\frac{4}{81}\frac{\Qbar!}{(\Qbar-4)!}$&$\frac{1}{2}(5-4\zeta_3)$\\
&&\\
\hline
&&\\
6(j)&$\frac{8}{81}\frac{\Qbar!}{(\Qbar-4)!}A$&$\frac52$\\
&&\\
\hline
&&\\
6(k)&$\frac{4}{81}\frac{\Qbar!}{(\Qbar-4)!}$&$\frac52$\\
&&\\
\hline
\end{tabular}
\caption{\label{4loopb}Four-loop results from Fig.~\ref{diagfourb}}
\end{center}
\end{table}

The simple pole contributions from the four-loop diagrams in Fig.~\ref{diagfour} were readily evaluated using standard techniques (see for instance Ref.~\cite{klein}). Those from Figs.~\ref{diagfoura}, \ref{diagfourb} may be extracted from Ref.~\cite{kaz}. The contributions from each four-loop diagram are listed in Tables~\ref{4loop}, \ref{4loopa} and \ref{4loopb} respectively, together with the corresponding symmetry factor. A factor of $g^4$ is understood in each case, and the $N$-dependent factor $C$ is given by
\be
C=\frac{1}{32}(N+30).
\ee
 When added and multiplied by a loop factor of 4, the leading and non-leading four-loop contributions to  $\gamma_{T_{\Qbar}}$ are found to be 
\be
\gamma^{(4)}_{T_{\Qbar}}=-\frac{1}{81}(g\Qbar)^4[42\Qbar+4(N-73)+2(6N+65)\zeta_3+5(N+30)\zeta_5],
\ee
once again in accord with the semiclassical results in Eqs.~\eqref{Delfull}, for general $g$.

\section{The large $g_*\Qbar$ calculation}\label{gQ}
In this section we discuss the large $g_*\Qbar$ limit of $\Delta_{T_{\Qbar}}$. The large $g_*\Qbar$ limit of $\Delta_{-1}$ as given by Eq.~\eqref{DelmQ} is readily obtained as
\be
\frac{\Delta_{-1}}{g_*}=\frac{3}{4g_*}\left[\frac34\left(\frac{4g_*\Qbar}{3}\right)^{\frac43}
+\frac12\left(\frac{4g_*\Qbar}{3}\right)^{\frac23}+\Ocal(1)\right].
\label{Delmlarge}
\ee
We follow the procedure described in Ref.~\cite{Bad2} for evaluating $\Delta_0$ by means of an approximation to the sum over $l$ followed by a numerical fit. The procedure involves selecting integers $N_1$, $N_2$ and picking $A\ge1$ such that $AR\mu_*$ is an integer (this represents a cut-off in the summation, beyond which we approximate it by an integral). The accuracy may be made as great as desired by increasing $N_1$, $N_2$ and $A$.  We obtain
\be
\Delta_0=\frac{N+8}{16}(R^2\mu^{*2}-1)^2\ln(AR\mu_*)+F(R\mu_*),
\label{largeDel}
\ee
where
\be
F(R\mu_*)=f_{N_2,A}(R\mu_*)-\frac14\sigma_{AR\mu_*}+\frac12\sum_{l=0}^{AR\mu_*}\sigma_l-\frac12\sum_{k=1}^{N_1}\frac{B_{2k}}{(2k)!}\sigma^{(2k-1)}_{AR\mu_*},
\label{Fdef}
\ee
and here
\begin{align}
f_{N_2,A}(R\mu_*)=&\frac12(AR\mu_*)^4\sum_{n=1,n\ne5}^{N_2}\frac{c_n}{(AR\mu_*)^{n-1}(n-5)}\nn
&+\frac{N+8}{16}(R^2\mu^{*2}-1)^2\left(\gamma-\frac32\right)-\frac{29}{24}(R^2\mu^{*2}-1)
-\left(\frac{11}{48}R^2\mu^{*2}-\frac{1}{10}\right)N.
\label{fdef}
\end{align}
With some help from one of the authors\cite{Cuomo} we have corrected some typos in the corresponding equations in Ref.~\cite{Bad2}, which were not reflected in their final results.
 The function $f_{N_2,A}(R\mu_*)$ derives from replacing the sum over $l$ for $l\ge AR\mu_*$ in Eq.~\eqref{lsum} by an integral over $l$. It is then appropriate to use the large $l$ expansion in Eq.~\eqref{largel}. The integral over $\frac{1}{l^{1+\epsilon}}$  corresponding to the $c_5$ term leads to a pole term in $\epsilon$. The potential pole in $\Delta_0$ is cancelled by the pole in the bare coupling, but the $O(\epsilon)$ term in $c_5$ in Eq.~\eqref{cvals} leads to the terms in the last line of Eq.~\eqref{fdef}. The details of the procedure may be found in Ref.~\cite{Bad2}. In Eq.~\eqref{Fdef}, we can set $d=4$. We now evaluate $F(R\mu_*)$ in Eq.~\eqref{Fdef} numerically. We take $N_1=4$, $N_2=10$ and $A=10$, using the same numbers as Ref.~\cite{Bad2} for comparison purposes. The result is then fitted with an expansion in $(R\mu_*)^{-2}$, starting from $(R\mu_*)^4$, with 4 parameters. We find that $F(R\mu_*)$ is given by
\begin{align}
F(R\mu_*)\sim& -(1.5559+0.2293N)(R\mu_*)^4+(1.8536+0.3231N)(R\mu_*)^2\nn
&-(0.4467+0.0826N)+\Ocal((R\mu_*)^{-2}),
\label{Fexp}
\end{align}
and this may be inserted into Eq.~\eqref{largeDel} to give the full result for $\Delta_0$.
Expanding $R\mu_*$ as given by Eq.~\eqref{mudefa} in terms of large $g_*\Qbar$, we find
\be
R\mu_*=\left(\frac{4g_*\Qbar}{3}\right)^{\frac13}+\frac13\left(\frac{4g_*\Qbar}{3}\right)^{-\frac13}+\ldots
\label{Rexp}
\ee
and then we obtain from Eq.~\eqref{largeDel}
\begin{align}
\Delta_0=&\left[\alpha+\frac{N+8}{48}\ln\left(\frac{4g_*\Qbar}{3}\right)\right]
\left(\frac{4g_*\Qbar}{3}\right)^{\frac43}\nn
&+\left[\beta-\frac{N+8}{72}\ln\left(\frac{4g_*\Qbar}{3}\right)\right]
\left(\frac{4g_*\Qbar}{3}\right)^{\frac23}+\Ocal(1),
\label{Delexp}
\end{align}
where 
\begin{align}
\alpha=&-0.4046-0.0854N,\nn
\beta=&-0.8218-0.0577N.
\label{alphs}
\end{align}
The results for $U(1)$ should be recovered by setting $N=2$; and indeed for $N=2$ we find Eqs.~\eqref{Fexp}, \eqref{Delexp}, \eqref{alphs} agree with the corresponding results given in Ref.~\cite{Bad2}.  

Following Ref.\cite{Bad2} and combining Eqs.~\eqref{gfix}, \eqref{Tscal}, \eqref{Delmlarge} and \eqref{Delexp}, we may write the full scaling dimension in the form
\begin{align}
\Delta_{T_{\Qbar}}=&\frac{1}{\epsilon}\left(\frac{4\epsilon \Qbar}{N+8}\right)^{\frac{d}{d-1}}\left[
\frac{3(N+8)}{16}+\epsilon\left(\alpha+\frac{3(3N+14)}{16(N+8)}\right)+\Ocal(\epsilon^2)\right]\nn
&+\frac{1}{\epsilon}\left(\frac{4\epsilon \Qbar}{N+8}\right)^{\frac{d-2}{d-1}}\left[
\frac{N+8}{8}+\epsilon\left(\beta-\frac{3N+14}{8(N+8)}\right)+\Ocal(\epsilon^2)\right]+\Ocal[(\epsilon\Qbar)^0]
\end{align}
In Ref.~\cite{JJ}, we found that we could reproduce the coefficients in the large $R\mu_*$ expansion of the $N$-dependent part of $\Delta_0$ (the terms involving $\omega^*_{++}$ and $\omega^*_{--}$ in Eq.~\eqref{ONDel}) by an analytic computation. This fails to work here; an analytic large-$R\mu_*$ expansion of  $\omega^*_{++}$ and $\omega^*_{--}$ as given by Eq.~\eqref{ompdef} leads to odd negative powers of $R\mu_*$, whereas our numeric computation in Eq.~\eqref{Fexp} only contains even powers of $R\mu_*$.  It appears that the simple properties of  $\omega^*_{++}$ and $\omega^*_{--}$ identified in Ref.~\cite{JJ}, in particular their expansion in powers of $\frac{J_l^2}{R^2\mu_*^2}$ , are not enough for our analytic computation to work in the $d=4$ case. A little trial and error indicates that the fact that in $d=3$, $n_l\propto\frac{d}{dl}J_l^2$, may also be crucial;  but further insight is required.

\section{Conclusions}

Approaches that extend the reach of (or even transcend the need for) perturbation theory have always been challenging, and are all the more interesting now because of the increased importance attached to multi-leg amplitudes, which can present formidable calculational obstacles at higher loop orders. In this paper we have followed Refs.~\cite{Bad2,rod,sann} in the application of semi-classical methods to  the calculation of $\phi^n$ amplitudes in $d=4$ renormalisable scalar theories with quartic interactions.  Ref~\cite{sann} generalises this calculation of Ref.~\cite{Bad2} from $U(1)$ to an $O(N)$  invariant interaction.  Another motivation for studying this class of theories is their (classical) scale invariance  (CSI). As remarked in Ref~\cite{sann}, the Standard Model (SM) is ``almost'' 
CSI. Indeed, in 1973, Coleman and Weinberg (CW)~\cite{CW}  had hoped to argue that the SM might indeed be viable with the omission of the Higgs (wrong-sign) $(\hbox{mass})^2$ term. This attractive idea failed. Neglecting Yukawa couplings (which seemed reasonable at the time) led to a Higgs mass prediction which was too small; but including the top quark Yukawa coupling destabilised the Higgs vacuum altogether \footnote{For a review of some controversy over this development, see Ref.~\cite{EJ}}. CW introduced the idea of {\it dimensional transmutation}\/ as a means of generating a physical mass scale in a CSI theory. The same phenomenon has  been pursued ~\cite{Einhorn:2014gfa,Einhorn:2015lzy,Einhorn:2016mws}
in the CSI form of quantum gravity~\cite{Stelle:1976gc,Fradkin:1981hx, 
Fradkin:1981iu, Avramidi:1985ki, Avramidi:1986zv, Avramidi:2000pia}. 

Our purpose here has been to compare the results of Ref~\cite{sann} with straightforward (albeit intricate) perturbation theory. Generally the results have supported the validity of the semi-classical  approximation, in its domain of validity. 

Future work might include the application of the semi-classical methods and perturbative methods used here to the remaining class of CSI theories with scalar self-interactions; that is $\phi^3$ theories in $d=6$; or even perhaps the case of  CSI  quantum gravity mentioned above.

\section*{Acknowledgements}

 We are grateful to Gabriel Cuomo for helpful correspondence. DRTJ thanks the Leverhulme Trust for the award
of an Emeritus Fellowship. This research was supported by the Leverhulme Trust, STFC
and by the University of Liverpool.


\begin{thebibliography}{1}


%\cite{Jack:2015tka}


\bibitem{wils1} K.G. Wilson, "Renormalization Group and Critical Phenomena. I. Renormalization Group and the Kadanoff Scaling Picture",  Phys. Rev. B4 (9): 3174–3183. 

\bibitem{wils2} K.G. Wilson,  "Renormalization Group and Critical Phenomena. II. Phase-Space Cell Analysis of Critical Behavior", Phys. Rev. B4 (9): 3184–3205.

\bibitem{wf}
K.G. Wilson and Michael E. Fisher, ``Critical exponents in 3.99 dimensions'',  Phys.Rev.Lett. 28 (1972) 240-243.


\bibitem{son} D.T. Son, ``Semiclassical approach for multiparticle production in
scalar theories", Nucl.Phys. B 477 (1996) 378-406. 
%hep-ph/9505338 
%doi:10.1016/0550-3213(96)00386-0


\bibitem{horm} S.~ Hellerman, D.~Orlando, S. Reffert and  M. Watanabe, ``On the CFT Operator Spectrum at Large Global Charge, JHEP \textbf{12} (2015) 071,  [arXiv:1505.01537 [hep-th]].


\bibitem{alos}L.~Alvarez-Gaume, O.~Loukas, D.~Orlando and S.~Reffert, ``Compensating strong coupling with large charge", JHEP \textbf{04} (2017) 059,
[arXiv:1610.04495 [hep-th]].

\bibitem{aos}L.~Alvarez-Gaume, D.~Orlando and S.~Reffert, ``Large charge at large N", JHEP \textbf{12} (2019) 142,
[arXiv:1909.12571 [hep-th]].


\bibitem{rod}
G.~Arias-Tamargo, D.~Rodriguez-Gomez and J.G.~Russo,
``The large charge limit of scalar field theories and the Wilson-Fisher fixed point at $\epsilon=0$,''
JHEP \textbf{10} (2019) 201,
%doi:10.1007/JHEP10(2019)201
[arXiv:1908.11347 [hep-th]].
%11 citations counted in INSPIRE as of 22 Jul 2020

\bibitem{Bad2} G.~Badel, G.~Cuomo, A.~Monin and R.~Rattazzi, 
``The epsilon expansion meets semiclassics", JHEP \textbf{11} (2019) 110 [arXiv:1909.01269[hep-th]].

\bibitem{sann}
O.~Antipin, J.~Bersini, F.~Sannino, Z.~Wang and C.~Zhang,``Charging the $O(N)$ model,''
Phys.Rev. D 102 (2020) 4, 045011, [arXiv:2003.13121 [hep-th]].

\bibitem{sann2}
O.~Antipin, J.~Bersini, F.~Sannino, Z.~Wang and C.~Zhang,``Charging the Walking 
$U(N)\otimes U(N)$ Higgs Theory as a Complex CFT,''
Phys. Rev. D 102, 12, 125033, [arXiv:2006.10078  [hep-th]].

\bibitem{alos2}L.~Alvarez-Gaume,  D.~Orlando and S.~Reffert, ``Selected Topics in the Large Quantum Number Expansion", arXiv:2008.03308 [hep-th].

\bibitem{Bad}
  G.~Badel, G.~Cuomo, A.~Monin and R.~Rattazzi,
 ``Feynman diagrams and the large charge expansion in $3-\varepsilon$ dimensions,''
  Phys.\ Lett.\ B {\bf 802} (2020) 135202
  %doi:10.1016/j.physletb.2020.135202
  [arXiv:1911.08505 [hep-th]].
  %%CITATION = doi:10.1016/j.physletb.2020.135202;%%
  %2 citations counted in INSPIRE as of 18 Feb 2020


 
 



\bibitem{rod2}Guillermo Arias-Tamargo, Diego Rodriguez-Gomez and Jorge G. Russo,
``On the UV completion of the $O(N)$ model in $6-\epsilon$ dimensions: a stable large-charge sector", 
 JHEP \textbf{09} (2020) 064  [arXiv:2003.13772 [hep-th]]


\bibitem{JJ}
 I.~Jack and D.R.T.~Jones,  
``Anomalous dimensions for $\phi^n$
in scale invariant $d=3$ theory.",  Phys.Rev. D 102 (2020) 8, 085012,  [arXiv:2007.07190 [hep-th]] 


\bibitem{Niel}
H.B.~Nielsen and S.~Chadha,
``On how to count Goldstone bosons,''
Nucl. Phys. B \textbf{105} (1976), 445-453
%doi:10.1016/0550-3213(76)90025-0

\bibitem{kaz}
D.~I.~Kazakov, O.~V.~Tarasov and A.~A.~Vladimirov,
``Calculation of Critical Exponents by Quantum Field Theory Methods,''
Sov. Phys. JETP \textbf{50} (1979), 521
JINR-E2-12249.
%127 citations counted in INSPIRE as of 23 Feb 2021

\bibitem{klein} 
  H.~Kleinert and V.~Schulte-Frohlinde,
  ``Critical properties of $\phi^4$-theories'', World Scientific (2001) .

\bibitem{Cuomo} G.~Cuomo, private communication.

\bibitem{CW}
S.R.~Coleman and  E.J. Weinberg, ``Radiative Corrections as the Origin of Spontaneous Symmetry Breaking'', Phys.Rev. D 7 (1973) 1888-1910.

\bibitem{EJ} M.B.~Einhorn and D.R.T.~Jones, ``The Effective potential, the renormalisation group and vacuum stability",  JHEP \textbf{04} (2007) 051,   hep-ph/0702295 [hep-ph]



\bibitem{Einhorn:2014gfa}
M.~B.~Einhorn and D.~R.~T.~Jones,
 ``Naturalness and Dimensional Transmutation in 
 Classically Scale-Invariant Gravity,''
 JHEP \textbf{1503} (2015) 047
 %doi:10.1007/JHEP03(2015)047
 [arXiv:1410.8513 [hep-th]].



\bibitem{Einhorn:2015lzy}
 M.~B.~Einhorn and D.~R.~T.~Jones,
 ``Induced Gravity I: Real Scalar Field,''
 JHEP \textbf {1601} (2016) 019
% doi:10.1007/JHEP01(2016)019
 [arXiv:1511.01481 [hep-th]].

\bibitem{Einhorn:2016mws}
 M.~B.~Einhorn and D.~R.~T.~Jones,
 ``Induced Gravity II: Grand Unification,''
 JHEP {\bf 1605} (2016) 185
 %doi:10.1007/JHEP05(2016)185
 [arXiv:1602.06290 [hep-th]]. 

%\cite{Stelle:1976gc}
\bibitem{Stelle:1976gc}
K.~S.~Stelle,
``Renormalization of Higher Derivative Quantum Gravity,''
Phys.\ Rev.\ D {\bf 16} (1977) 953.
%%CITATION = PHRVA,D16,953;%%

%\cite{Fradkin:1981hx}
\bibitem{Fradkin:1981hx}
  E.~S.~Fradkin and A.~A.~Tseytlin,
  ``Renormalizable Asymptotically Free Quantum Theory of Gravity,''
  Phys.\ Lett.\ B {\bf 104} (1981) 377.
  %%CITATION = PHLTA,B104,377;
  
%\cite{Fradkin:1981iu}
\bibitem{Fradkin:1981iu}
  E.~S.~Fradkin and A.~A.~Tseytlin,
  ``Renormalizable asymptotically free quantum theory of gravity,''
  Nucl.\ Phys.\ B {\bf 201} (1982) 469.
  %%CITATION = NUPHA,B201,469;%%

%\cite{Avramidi:1985ki}
\bibitem{Avramidi:1985ki}
I.~G.~Avramidi and A.~O.~Barvinsky,
``Asymptotic Freedom In Higher Derivative Quantum Gravity,''
Phys.\ Lett.\ B {\bf 159} (1985) 269.

%\cite{Avramidi:1986zv}
\bibitem{Avramidi:1986zv}
  I.~G.~Avramidi,
  ``Asymptotic Behavior of the Quantum Theory of Gravity With Higher Order Derivatives,''
  Sov. J. Nucl. Phys. 44 (1986) 160.
%(Yad.\ Fiz.\  {\bf 44} (1986) 255)
  %%CITATION = YAFIA,44,255;%%

%\cite{Avramidi:2000pia}
\bibitem{Avramidi:2000pia}
I.~G.~Avramidi,
  ``Heat kernel and quantum gravity,'' Lect.\ Notes Phys.\ M {\bf 64} (2000) 1.
%This is essentially a reproduction of the author's 1986 PhD thesis [hep-th/9510140].
  %%CITATION = LNPHA,M64,1;%%

\end{thebibliography}
\end{document}